\documentclass[12pt,preprint]{aastex}

\shorttitle{Search for Ultra High-Energy Neutrinos with AMANDA-II}
\shortauthors{Ackermann et al.}

\begin{document}

\title{Search for Ultra High-Energy Neutrinos with AMANDA-II}

\author{
IceCube Collaboration: 
M.~Ackermann\altaffilmark{33},
J.~Adams\altaffilmark{11},
J.~Ahrens\altaffilmark{22},
K.~Andeen\altaffilmark{21},
J.~Auffenberg\altaffilmark{32},
X.~Bai\altaffilmark{24},
B.~Baret\altaffilmark{9},
S.~W.~Barwick\altaffilmark{16},
R.~Bay\altaffilmark{5},
K.~Beattie\altaffilmark{7}
% ...
}
% ...
\altaffiltext{5}{Dept.~of Physics, University of California, Berkeley, CA 94720, USA}
% ...
\altaffiltext{7}{Lawrence Berkeley National Laboratory, Berkeley, CA 94720, USA}
% ...
\altaffiltext{9}{Vrije Universiteit Brussel, Dienst ELEM, B-1050 Brussels, Belgium}
% ...
\altaffiltext{11}{Dept.~of Physics and Astronomy, University of Canterbury, Private Bag 4800, Christchurch, New Zealand}
% ...
\altaffiltext{16}{Dept.~of Physics and Astronomy, University of California, Irvine, CA 92697, USA}
% ...
\altaffiltext{21}{Dept.~of Physics, University of Wisconsin, Madison, WI 53706, USA}
\altaffiltext{22}{Institute of Physics, University of Mainz, Staudinger Weg 7, D-55099 Mainz, Germany}
% ...
\altaffiltext{24}{Bartol Research Institute and Department of Physics and Astronomy, University of Delaware, Newark, DE 19716, USA}
% ...
\altaffiltext{32}{Dept.~of Physics, University of Wuppertal, D-42119 Wuppertal, Germany}
\altaffiltext{33}{DESY, D-15735 Zeuthen, Germany}
% ...




\begin{abstract}
A search for diffuse neutrinos with energies in excess of 10$^{5}$~GeV is conducted 
with AMANDA-II data recorded between 2000 and 2002. Above 10$^{7}$~GeV, the Earth is 
essentially opaque to neutrinos. This fact, combined with the limited overburden of the 
AMANDA-II detector (roughly 1.5 km), concentrates these ultra high-energy neutrinos at 
the horizon. The primary background for this analysis is bundles of downgoing, high-energy 
muons from the interaction of cosmic rays in the atmosphere. No statistically significant 
excess above the expected background is seen in the data, and an upper limit is set on 
the diffuse all-flavor neutrino flux of E$^{2}$ $\Phi$$_{\mathrm{90\% CL}}$ $<$ 2.7 
$\times$ 10$^{-7}$ GeV cm$^{-2}$ s$^{-1}$ sr$^{-1}$ valid over the energy range of 2 
$\times$ 
10$^{5}$~GeV to 10$^{9}$~GeV. A number of models which predict neutrino fluxes from 
active galactic nuclei are excluded at the 90\% confidence level.
\end{abstract}


\section{Introduction}
AMANDA-II (Antarctic Muon and Neutrino Detector Array), a neutrino telescope at the 
geographical South Pole designed to detect Cherenkov light from secondary particles 
produced in collisions between neutrinos and Antarctic ice, has placed limits on the flux 
from point-like and diffuse sources of astrophysical neutrinos \citep{ach07, ack04, 
ack05, ahr03a, ahr03b}. This work describes a search for neutrinos with energies above 
10$^{5}$~GeV, which we define as ultra high-energy (UHE) neutrinos. These neutrinos are 
of interest because they could be associated with the potential acceleration of hadrons 
by active galactic nuclei \citep{man95, mpr00, hal97, pro96, ste92}; they could 
potentially be produced by exotic phenomena such as the decay of topological defects 
\citep{sig98} or possibly associated with the Z-burst mechanism \citep{yos98}; and they 
are guaranteed by-products of the interactions of high-energy cosmic rays with the cosmic 
microwave background \citep{eng01}. 

\begin{figure}
\epsscale{.80}
\plotone{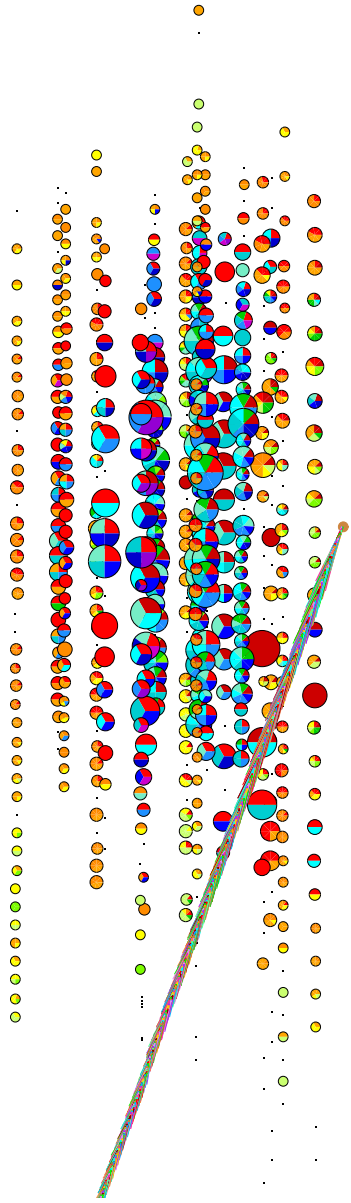}
\caption{Simulated muon neutrino event with an energy of 3.8 $\times$ 10$^{8}$~GeV. The 
muon passes roughly 70~m outside the instrumented volume of the detector. Colored circles 
represent hit OMs. The color of the circle indicates the hit time (red is earliest), with 
multiple colors indicating multiple hits in that OM. The size of the circle is correlated 
with the number of photons produced.\label{figevt}} 
\end{figure}

\begin{figure}
\epsscale{.80}
\plotone{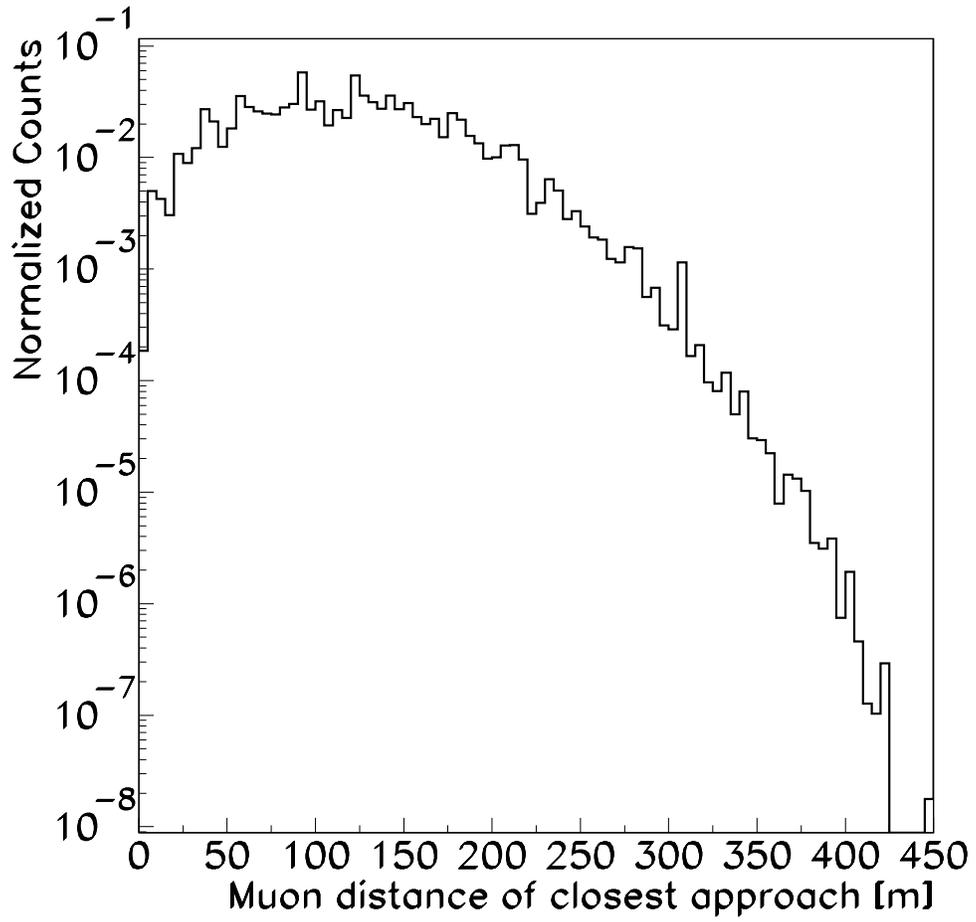}
\caption{Distance of closest approach to the detector center for muons from UHE muon 
neutrinos (shown with an E$^{-2}$ spectrum) which pass all selection criteria of this 
analysis.\label{figdist}} 
\end{figure}

This analysis is sensitive to all three flavors of neutrinos. Leptons and cascades from 
UHE electron, muon and tau neutrinos create bright, energetic events (Fig. \ref{figevt}) 
which can be identified by AMANDA-II as far as 450~m from the center of the array (Fig. 
\ref{figdist}). The sensitivity of this analysis starts at energies roughly coincident 
with the highest energy threshold of other diffuse analyses conducted with AMANDA-II 
\citep{ach07, ack04}.

At UHE energies, the interaction length of neutrinos in rock is shorter than the diameter 
of the Earth \citep{gan98}, so neutrinos from the Northern Hemisphere will interact 
before reaching AMANDA-II. Combined with the limited overburden above the AMANDA-II 
detector, this concentrates UHE events at the horizon. This contrasts with the majority 
of other astrophysical neutrino analyses completed using data from the AMANDA-II detector, 
which search for neutrinos from the Northern Hemisphere with energies below 10$^{5}$~GeV. 

The flux of atmospheric neutrinos is negligible at UHE energies, with fewer than 10 events 
in three years expected from the model in \citet{lip93} after intermediate UHE selection 
criteria have been applied. This drops to 0.1 events after application of all selection 
criteria. Similarly, there are fewer than 0.6 events expected in three years at the final 
selection level from prompt neutrinos from the decay of charmed particles produced in the 
atmosphere (using the ``C'' model from \citet{zas93}). Therefore, the primary background 
for the UHE analysis is composed of many lower energy processes that mimic higher energy 
signal events. Cosmic ray collisions in the upper atmosphere that generate large numbers 
of nearly parallel muons (or ``muon bundles'') can generate high-energy signatures even 
though the individual muons have much lower energy than single leptons or cascades from 
UHE neutrinos. 
Signal and background events spread light over roughly equivalent areas in the detector, 
but UHE neutrino events are distinguishable because they have higher energy and higher 
light density than background events. Specialized selection criteria which use these 
properties, as well as differences in reconstruction variables, separate the UHE neutrinos 
from the background of muon bundles from atmospheric cosmic rays.

\begin{figure}
\plotone{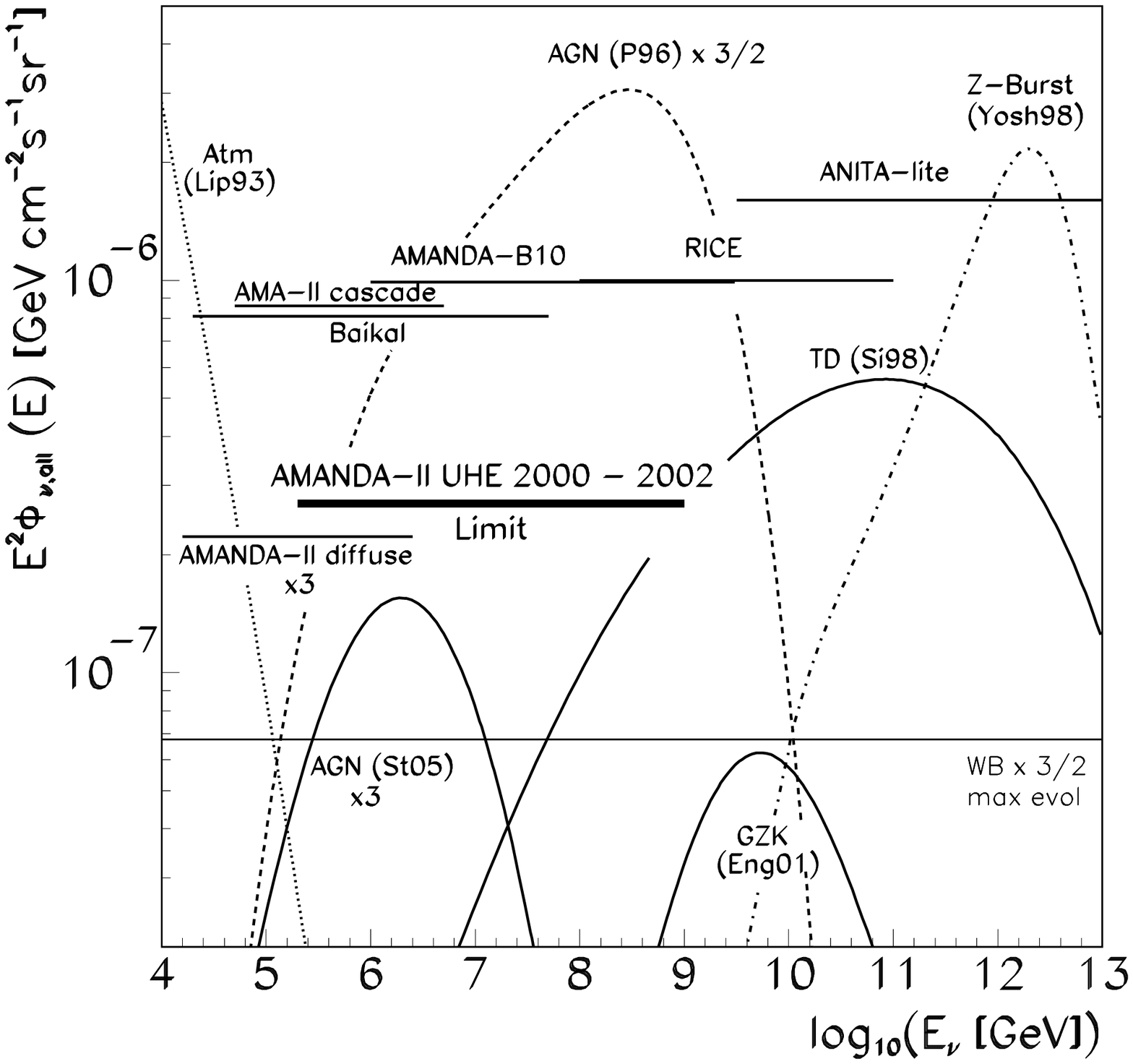}
\caption{All-flavor UHE neutrino flux limit for 2000-2002 over the range which contains 
the central 90\% of the expected signal with an E$^{-2}$ spectrum. Also shown are several 
representative models:  St05 from \citet{ste05} multiplied by 3, P96 from \citet{pro96} 
multiplied by 3/2, Eng01 from \citet{eng01}, Si98 from \citet{sig98}, Yosh98 from 
\citet{yos98}, Lip93 from \citet{lip93}, and the Waxman-Bahcall upper bound \citep{bah98} 
multiplied by 3/2. Existing experimental limits shown are from the RICE \citep{kra06}, 
ANITA-lite \citep{bar06}, and Baikal \citep{ayn06} experiments, the UHE limit from 
AMANDA-B10 \citep{ack05}, the lower-energy diffuse muon limit multiplied by 3 
\citep{ach07} and cascade limit \citep{ack04} from AMANDA-II.\label{figrep}} 
\end{figure}

Limits have been placed on the all-flavor neutrino flux in the ultra high-energy range by 
other experiments (Fig. \ref{figrep}). Additionally, a previous analysis using an earlier 
configuration of the AMANDA detector called AMANDA-B10, consisting of 302 optical modules 
\citep{ack05}, has placed limits on the all-flavor UHE neutrino flux (Fig. \ref{figrep}). 
This analysis uses 677 optical modules (OMs) of the AMANDA-II detector and gives a 
combined result using data from three years (2000-2002) with a livetime of 456.8 days.

A description of the AMANDA-II detector is given in section 2. Sections 3 and 4 discuss 
possible sources of astrophysical neutrinos and background, and the simulation of both. 
The selection criteria used to separate UHE neutrino signals from background are discussed 
in section  5. A study of systematic uncertainties is presented in section 6, and the 
results are shown in section 7.

\section{The AMANDA-II Detector}

The AMANDA-II detector \citep{ahr04a} consists of 677 OMs stationed between 1500 m and 
2000 m beneath the surface of the Antarctic ice at the geographic South Pole. The OMs are 
deployed on nineteen vertical cables (called ``strings'') arranged in three roughly 
concentric circles, giving the detector a cylindrical shape with a diameter of 
approximately 200 m. 

Each OM contains a Hamamatsu 8-inch photomultiplier tube (PMT) coupled with silicon gel to a 
spherical glass pressure housing for continuity of the index of refraction. The OMs are 
connected to the surface by cables which supply high voltage and carry the signal from 
the PMT to data acquisition electronics at the surface. The inner ten strings 
use electrical analog signal transmission, while the outer nine strings primarily use 
optical fiber transmission \citep{ahr04a}. 

The AMANDA-II detector uses a majority trigger of 24 OMs recording a voltage above a set 
threshold (a ``hit'') within a time window of 2.5 ${\mu}$s. An OM records the maximum 
amplitude, as well as the leading edge time and time over threshold for each hit, with 
each OM recording a maximum of eight hits per event. Each photoelectron has approximately 
a 3\% chance of producing an afterpulse caused by ionization of residual gas inside the 
PMT \citep{ham99}. This afterpulse follows several ${\mu}$s after the generating hit and 
aids in the detection of UHE events.

AMANDA-II has been collecting data since February 2000. In 2002/2003, waveform digitizers 
were installed which record the full pulse shape from each OM \citep{sil05}. In 2005 
deployment began on IceCube \citep{ahr04b}, a 1 km$^{3}$ array of digital OMs which now 
encompasses the AMANDA-II detector.

\section{Astrophysical Neutrino and Background Sources}
Astrophysical neutrinos with energies in excess of 10$^{5}$~GeV may be produced by a 
variety of sources. A number of theories predict neutrino fluxes from active galactic 
nuclei (AGN) peaking near 10$^{6}$~GeV. In these scenarios, protons are accelerated by the 
first order Fermi mechanism in shock fronts. In the favored mechanism for neutrino 
production, these protons interact with the ambient photon field either in the cores 
\citep{ste92} or jets \citep{pro96,hal97,mpr00,man95} of the AGN and produce neutrinos 
via the process:
\begin{equation}
p + \gamma \rightarrow \Delta^{+} \rightarrow \pi^{+} [+ n] \rightarrow \nu_{\mu} + \mu^{+} 
\rightarrow \nu_{\mu} + e^{+} + \nu_{e} + \overline{\nu_{\mu}},\\
\end{equation}
\noindent
resulting in a $\nu_{e}$:$\nu_{\mu}$:$\nu_{\tau}$ flavor ratio of 1:2:0 at the 
source\footnote{Neutrino flavor oscillation changes the flavor ratio to 1:1:1 at the 
Earth. See \citet{kas05} for a discussion of different flavor ratios.}. The energy 
spectrum of the neutrinos produced by these interactions generally follows the E$^{-2}$ 
spectrum of the protons. Theoretical bounds can be placed on the flux of these neutrinos 
based on the observation of cosmic rays if the p-$\gamma$ reaction takes place in the jet 
or other optically thin region of the AGN \citep{bah98, mpr00}. 

UHE neutrinos are also associated with models created to explain the apparent excess of 
cosmic rays at the highest energies. One scenario involves the decay of massive objects, 
such as topological defects created by symmetry breaking in the early universe 
\citep{sig98}. These objects decay close to the Earth into showers of particles, 
eventually producing neutrinos as well as a fraction of the highest-energy cosmic rays.
Z-burst models could also produce some of the highest-energy cosmic rays through the 
interaction of neutrinos with energies in excess of 10$^{13}$~GeV with relic neutrinos 
via the Z$^{0}$ resonance. Since these neutrino-neutrino interactions are rare, it is 
possible to directly search for the UHE neutrino fluxes required by this mechanism 
\citep{yos98,kal02a}. It should be noted that Z-burst scenarios which predict the highest 
flux of neutrinos have already been eliminated by previous experiments \citep{bar06}. 
Additionally, Z-burst models predict fluxes of neutrinos which peak at energies above the 
sensitivity of this analysis or require unrealistic assumptions and are mentioned 
primarily for completeness.

A guaranteed source of UHE neutrinos comes from the interaction of high-energy cosmic 
rays with the cosmic microwave background (see e.g. \citet{eng01} and \citet{kal02b}). 
However, the flux predictions of these GZK neutrinos are generally several orders of 
magnitude lower than most of the fluxes listed previously.

The background for this analysis consists of bundles of muons from cosmic rays. The cosmic 
rays follow an E$^{-2.7}$ spectrum until about 10$^{6}$~GeV, where the flux steepens to 
E$^{-3}$ \citep{hor03}. They come only from the Southern Hemisphere because bundles from 
other directions are absorbed by the Earth. According to simulations, there can be as 
many as 20,000 muons in one bundle spread over a rms cross-sectional area as large as 
200~m$^{2}$, and the highest-energy events can deposit energies as large as 2.4 $\times$ 
10$^{6}$~GeV in the ice around the AMANDA-II detector. 


\section{Simulation and Experimental Data}
UHE neutrinos are simulated using the All Neutrino Interaction Simulation (ANIS) package 
\citep{kow05} to generate and propagate the neutrinos through the Earth. All three flavors of 
neutrinos are simulated with energies between 10$^{3}$~GeV and 10$^{12}$~GeV. The 
resulting muons and taus are propagated through the rock and ice near the detector using 
the Muon Monte Carlo (MMC) simulation package \citep{chi04}. Finally, the detector 
response is simulated using the AMASIM2 simulation package \citep{hun98}.

The background muon bundles from cosmic rays are generated using the CORSIKA simulation 
program with the QGSJET01 hadronic interaction model \citep{hec99}. At early levels of 
this analysis, cosmic ray primaries are generated with composition and spectral indices 
from \citet{wie99}, with energies of the primary particles ranging between 8 $\times$ 
10$^{2}$~GeV and 10$^{11}$~GeV. At later levels of this analysis, the lower energy 
primaries have been removed by the selection criteria, and a new simulated data set is 
used with energy, spectral shape, and composition optimized to simulate high-energy 
cosmic rays more efficiently. In this optimized simulation, the energy threshold is 
raised to 8 $\times$ 10$^{4}$~GeV and only proton and iron primaries are generated with a 
spectrum of E$^{-2}$. These primaries are reweighted following the method outlined in 
\citet{gla99}. This optimized simulation is used for level 2 of the analysis and beyond 
(see Table \ref{tbl-cuts}). For 2001 and 2002, the background simulation is further 
supplemented with the inclusion of a third set of simulated data with the energy 
threshold increased to 10$^{6}$~GeV. For all sets of background simulation, the resulting 
particles are propagated through the ice using MMC, and the detector response is 
simulated using AMASIM2.


Data used in this analysis were recorded in the time period between February 2000 and 
November 2002, with breaks each austral summer for detector maintenance, engineering, 
and calibration lasting approximately four months. In addition to maintenance downtime, 
the detector also has a brief period while recording each event in which it cannot record 
new events. Runs with anomalous characteristics (such as excessive trigger rates or large 
numbers of OMs not functioning) are discarded and a method which removes non-physical 
events caused by short term detector instabilities is applied \citep{poh04}. These 
factors combine to give a deadtime of 17\% of the data taking time for 2000, 22\% of the 
data-taking time for 2001, and 15\% of the data taking time for 2002. Additionally, 26 
days are excluded from 2000 because the UHE filtered events are polluted with high number 
of events with incomplete hit information, likely due to a minor detector malfunction. 
Taking these factors into account, there are 173.5 days of livetime in 2000, 192.5 days 
of livetime in 2001, and 205.0 days of livetime in 2002. Finally, 20\% of the data from 
each year is set aside for comparison with simulations and to aid in the determination of 
selection criteria, leading to a total livetime for the three years of 456.8 days.

\section{Analysis}
Twenty percent of the data from 2000 to 2002 (randomly selected from throughout the three 
years) is used to test the agreement between background simulations and observations. 
In order to avoid biasing the determination of selection criteria, this 20\% is then 
discarded, and the developed selection criteria are applied to the remaining 80\% of 
the data. A previous UHE analysis was performed on only the 2000 data using different 
selection criteria than those described below (see \citet{ger05} and \citet{ger06} for a 
more detailed description). 
For 2001-2002, improved reconstruction techniques such as cascade reconstructions 
\citep{ahr03c} were added to the analysis, and the new selection criteria described below 
were devised in a blind manner. These selection criteria were also applied to the 
2000 data to derive a combined three year limit. Due to differences in hit selection for 
reconstruction between 2000 and 2001-2002, the E$^{-2}$ signal passing rate at the 
final selection level for the year 2000 is approximately 60\% of the rate for the years 
2001 and 2002.

In order to maximize the limit setting potential, the selection criteria are initially 
determined by optimizing the model rejection factor \citep{hil03} given by
\begin{equation}
\mathrm{MRF} = \frac{\mathrm{\bar{\mu}}_{90}}{\mathrm{N}_{\mathrm{signal}}},
\end{equation}
where $\bar{\mu}_{90}$ is 90\% confidence level (CL) average event upper limit given by 
\citet{fel98}, and N$_{\mathrm{signal}}$ is the number of muon neutrinos expected for 
the signal being tested, in this case an E$^{-2}$ flux. 
The selection criteria for this analysis are summarized in Table 
\ref{tbl-cuts} and described below.

\begin{figure}
\plotone{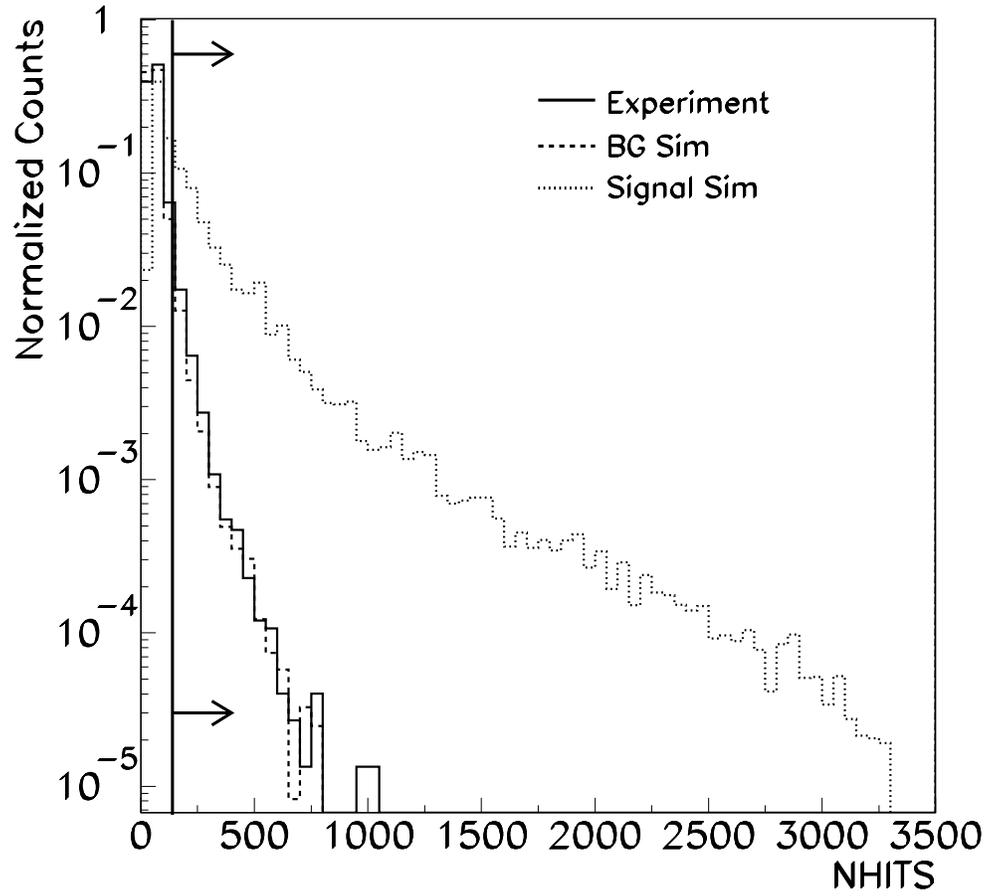}
\caption{NHITS distribution for the experiment, background, and E$^{-2}$ muon neutrino 
signal simulations before level 1 of this analysis.\label{fignhits}}
\end{figure}

\begin{figure}
\plotone{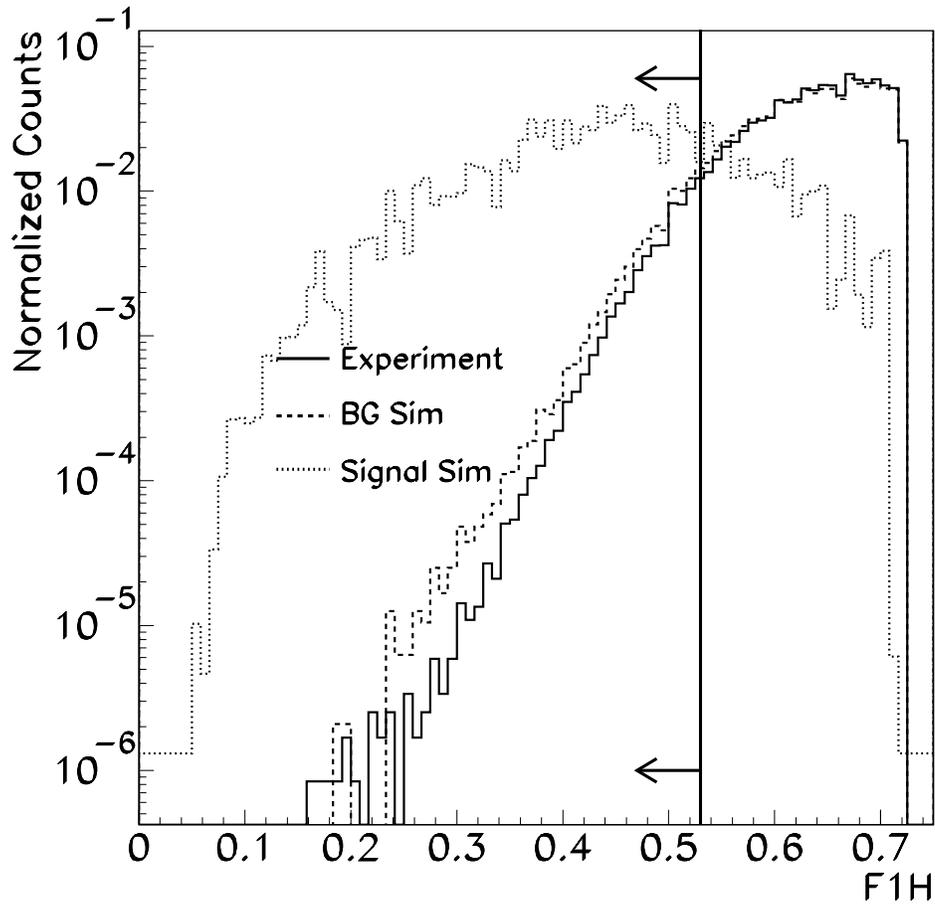}
\caption{Distribution of F1H (the fraction of OMs with a single hit) for the experiment, 
background, and E$^{-2}$ muon neutrino signal simulations after level 1 of this analysis. 
The average F1H drops with energy (see Fig. \ref{figf1he}).\label{figf1h}}
\end{figure}

\begin{figure}
\plotone{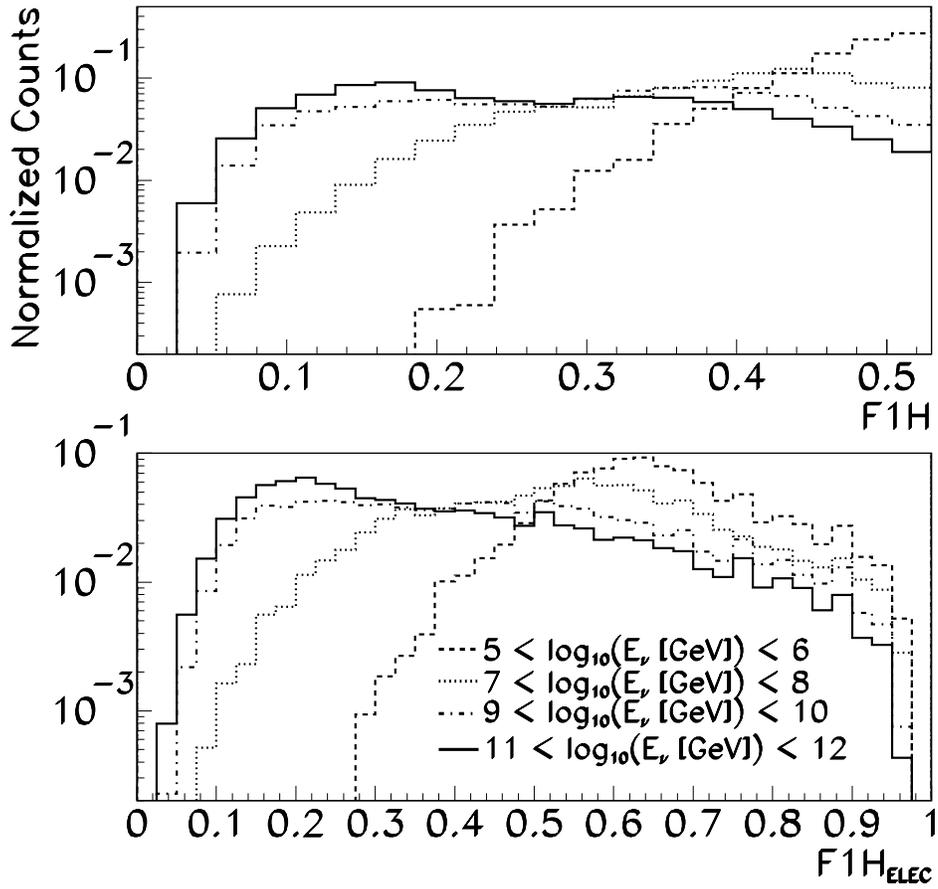}
\caption{F1H (top) and F1H$_{\mathrm{ELEC}}$ (bottom) distributions for various energy 
decades of muon neutrino signal. These variables serve as rough estimator of energy for 
the UHE analysis.\label{figf1he}}
\end{figure}

This analysis exploits the differences in total energy and light deposition between 
bundles of many low-energy muons and single UHE muons or cascades from UHE neutrinos.  
UHE neutrinos deposit equal or greater amounts of light in the ice than background muon 
bundles. In addition to being lower energy, background muon bundles spread their light 
over the cross sectional area of the entire muon bundle, rather than just along a single 
muon track or into a single cascade. Both signal and background events can have a large 
number of hits in the array, but for the same number of hit OMs, the muon bundle has a 
lower total number of hits, NHITS (recall each OM may have multiple separate hits in one 
event; see Fig. \ref{fignhits}). The number of hits for UHE neutrinos is increased by the 
tendency of bright signals to produce afterpulses in the PMT. Background muon bundles 
also have a higher fraction of OMs with a single hit (F1H), while a UHE neutrino 
generates more multiple hits (Fig. \ref {figf1h}). The F1H variable is correlated with 
energy (Fig. \ref{figf1he}) and is effective at removing lower energy background muon 
bundle events. The level 1 and 2 selection criteria require that NHITS $>$ 140 and F1H 
$<$ 0.53 and reduce the background by a factor of 2 $\times$ 10$^{3}$ relative to trigger 
level (level 0 on Tables \ref{tbl-passc} and \ref{tbl-passm}). 

At this point the data sample is sufficiently reduced that computationally intensive 
reconstructions become feasible. Reconstruction algorithms used in this analysis employ a 
maximum likelihood method which takes into account the absorption and scattering of light 
in ice. For muons, the reconstruction compares time residuals to those expected from a 
Cherenkov cone for a minimally ionizing muon \citep{ahr04a}, while the cascade 
reconstruction uses Cherenkov light from an electromagnetic cascade for comparison 
\citep{ahr03c}. Reconstructions which are optimized for spherical (cascade) depositions 
of light are used to distinguish UHE neutrinos from background muon bundles which happen 
to have a large energy deposition, such as a bremsstrahlung or e$^{+}$ e$^{-}$ pair 
creation, inside the detector fiducial volume. 

\begin{figure}
\plotone{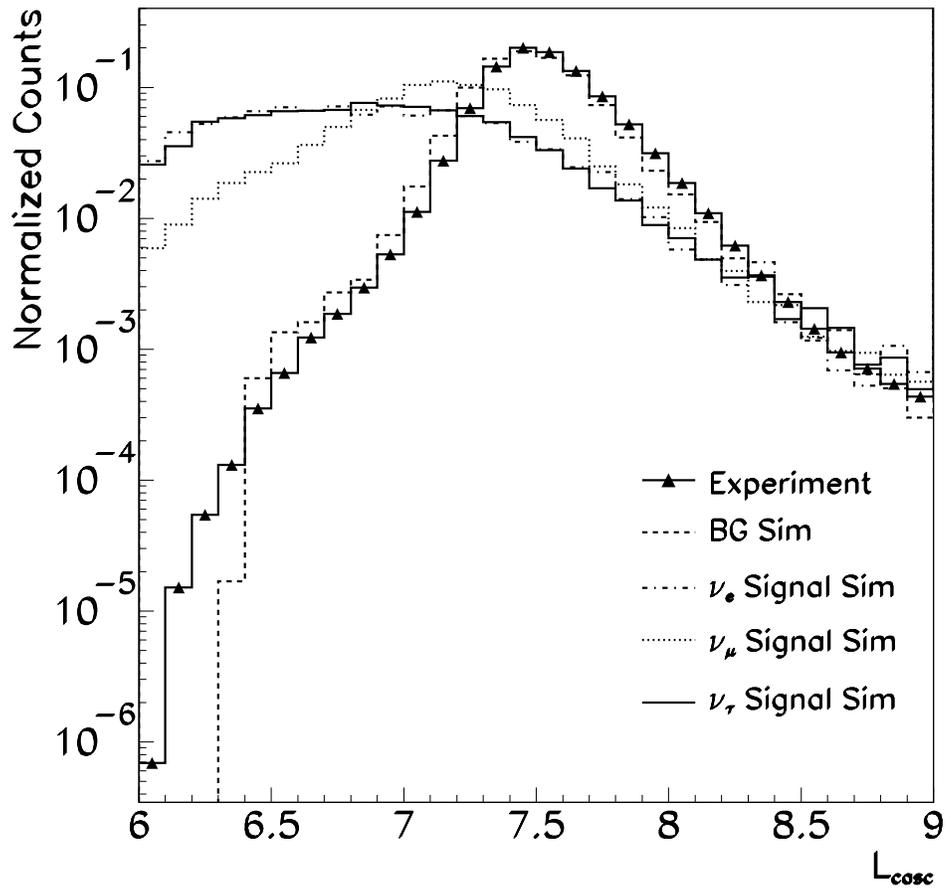}
\caption{Distribution of L$_{\mathrm{casc}}$ for the experiment, background, and 
E$^{-2}$ electron, muon, and tau neutrino signal simulations after level 2 of this 
analysis. Events with L$_{\mathrm{casc}} < 7$ are ``cascade-like,'' and events with 
L$_{\mathrm{casc}} \ge 7$ are ``muon-like.''\label{figjk23}} 
\end{figure}

Before application of the level 3 selection criteria, the data sets are split into 
``cascade-like'' and ``muon-like'' subsets. This selection is performed using the 
negative log likelihood of the cascade reconstruction (L$_\mathrm{{casc}}$, see Fig. 
\ref{figjk23}), where events with a L$_\mathrm{{casc}} < 7$ are considered 
``cascade-like.'' 

\subsection{``Cascade-like'' Events}
\begin{figure}
\plotone{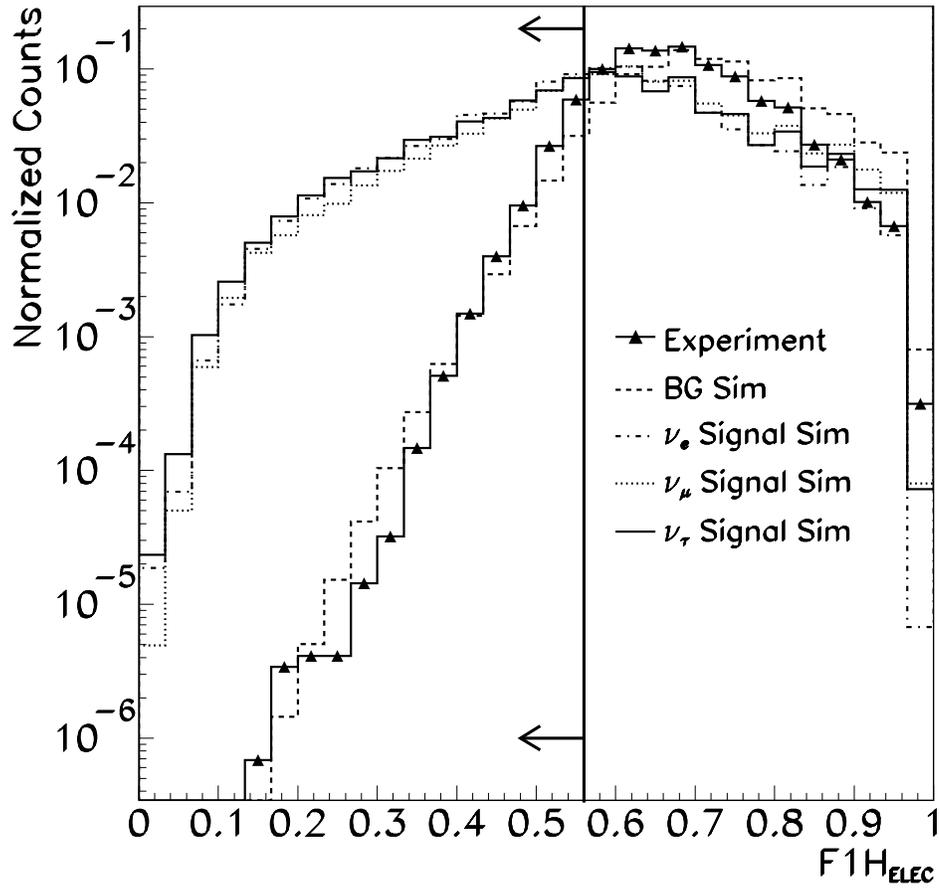}
\caption{F1H$_{\mathrm{ELEC}}$ (the fraction of electrical OMs with a single hit) 
distribution for the experiment, background, and E$^{-2}$ electron, muon, and tau 
neutrino signal simulations in the ``cascade-like'' subset after level two of this 
analysis.\label{figf1helec}}
\end{figure}

\begin{figure}
\plotone{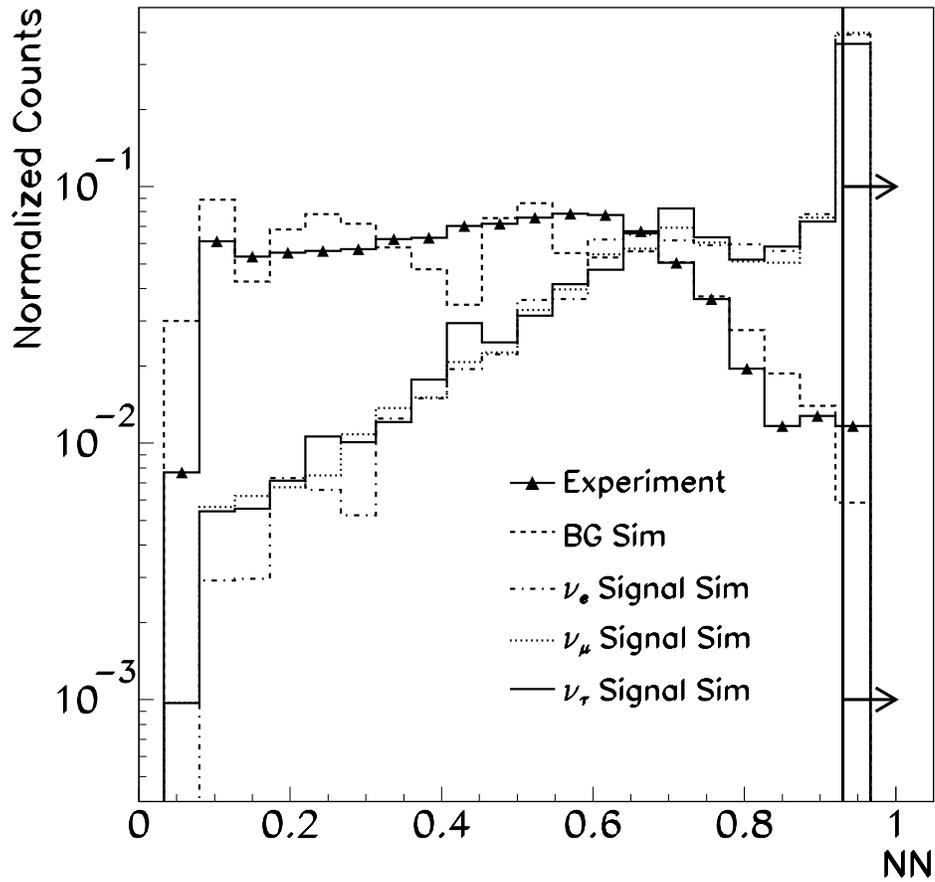}
\caption{Distribution of neural net output for the experiment, background, and E$^{-2}$ 
electron, muon, and tau neutrino signal simulations in the ``cascade-like'' subset after 
level two of this analysis. Signal events are expected near one, while background events 
are expected near zero.\label{fignn}}
\end{figure}

Background events in the ``cascade-like'' subset are characterized by either a large light 
deposition in or very near the instrumented volume of AMANDA-II or a path which clips the 
top or bottom of the array. In either case, the energy deposition is significantly less 
than the energy deposited by a UHE neutrino, allowing application of selection 
criteria which correlate with energy. One of these is F1H$_{\mathrm{ELEC}}$ (Fig. 
\ref{figf1helec}), a variable similar to the F1H variable described above, except that it 
uses only OMs whose signal is brought to the surface by electrical cables. The signal 
spreads as it propagates up the cable, causing hits close together in time to be 
combined. This gives F1H$_{\mathrm{ELEC}}$ a different distribution from F1H, and both 
variables are good estimators of energy deposited inside the detector (Fig. 
\ref{figf1he}). Additionally, the fraction of OMs with exactly four hits (F4H) is another 
useful energy indicator. The value of four hits was chosen as a compromise between the 
number of hits expected from OMs with electrical cables and OMs with optical 
fibers. OMs with optical fibers typically have more hits than OMs with electrical 
cables because very little pulse spreading occurs as the signal propagates up 
the fiber. The level 3 selection criteria uses the output of a neural net with 
F1H$_{\mathrm{ELEC}}$, F4H, and F1H as input variables (Fig. \ref{fignn}). As selection 
levels 4 and 5, separate applications of the F4H and F1H$_{\mathrm{ELEC}}$ variables 
remove persistent lower energy background events. 

The remaining background muon bundles have a different hit distribution than UHE 
neutrinos. In the background muon bundles, a large light deposition can be washed out by 
the continuous, dimmer light deposition from hundreds to tens of thousands of muons 
tracks. In contrast, UHE muons can have one light deposition that is several orders of 
magnitude brighter than the light from the rest of the muon track and looks very similar 
to bright cascades from UHE electron and tau neutrinos. For all cases, the initial cascade 
reconstruction is generally concentric with this large energy deposition, so ignoring OMs 
that are within 60 m of the initial cascade reconstruction reduces the fraction of OMs 
that are triggered with photons from the cascade. For background, the remaining light 
will be dominated by light depositions from the tracks of the muon bundles and be less 
likely to reconstruct as a cascade. In contrast, signal events, with their energetic 
cascades, will still appear cascade-like and reconstruct with a better likelihood 
(L$_{60}$). The final selection criteria for ``cascade-like'' events (chosen by 
optimizing the MRF) requires that these events be well reconstructed by a cascade 
reconstruction performed using only OMs with distances greater than 60~m and reduces the 
background expectation to 0 events for this subset.

The number of events at each selection level for experiment, background, and signal 
simulation for the ``cascade-like'' subset are shown in Table \ref{tbl-passc}. 
 
\subsection{``Muon-like'' Events}
\begin{figure}
\plotone{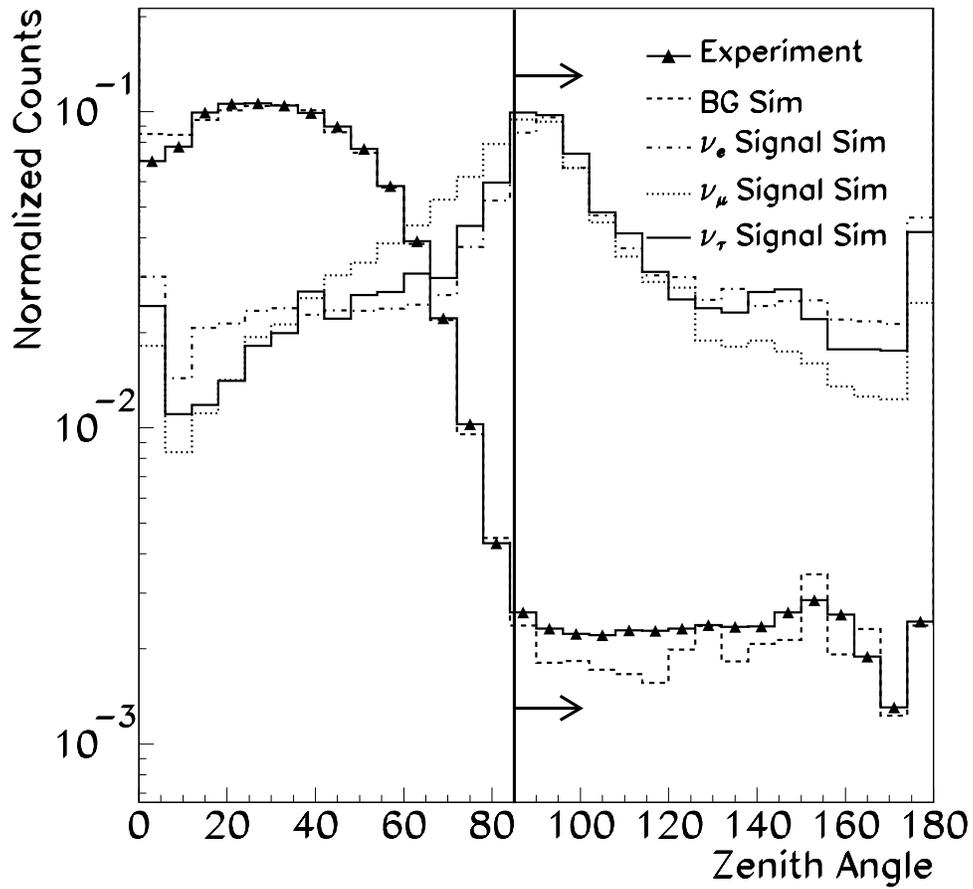}
\caption{Reconstructed zenith angle distribution for the experiment, background, and 
E$^{-2}$ electron, muon, and tau neutrino signal simulations in the ``muon-like'' subset 
after level two of this analysis. Zenith angles of 90$^{\circ}$ correspond to horizontal 
events, and zenith angles of 0$^{\circ}$ are downgoing events.\label{figzen}}
\end{figure}

\begin{figure}
\plotone{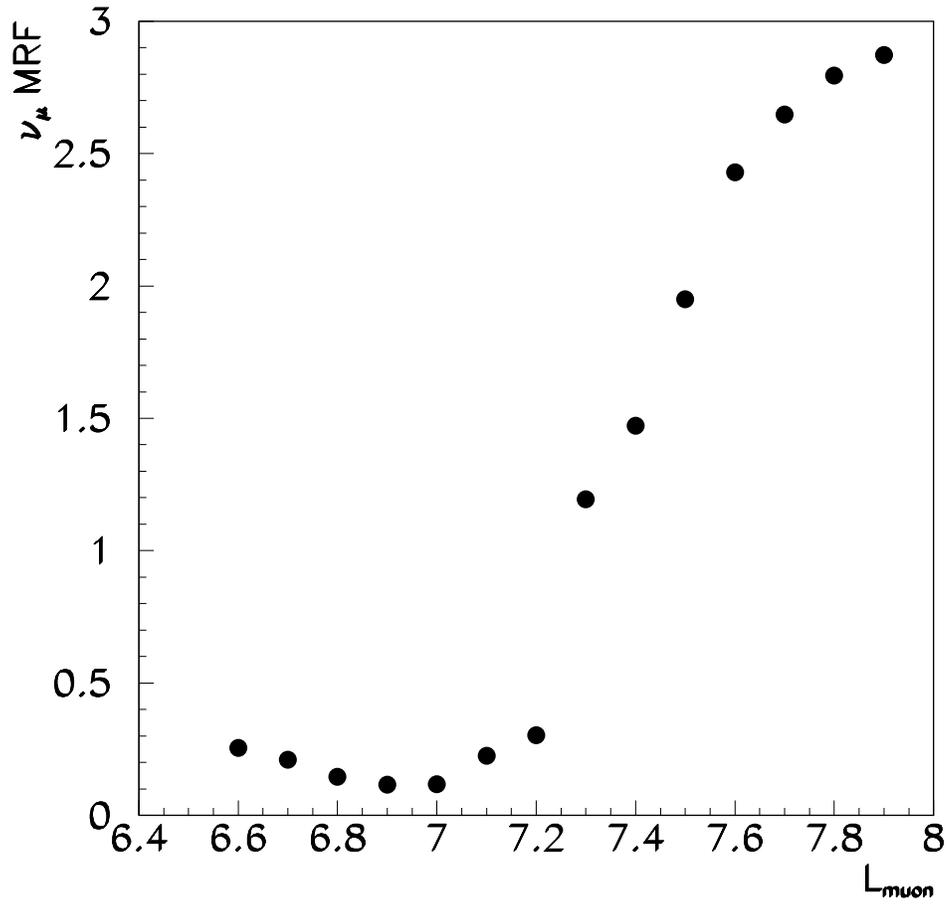}
\caption{Model rejection factor for 10$^{-6}$ $\times$ E$^{-2}$ muon neutrinos in the 
``muon-like'' subset as a function of cut level for L$_{\mathrm{muon}}$.\label{figmrf}} 
\end{figure}

Background events in the ``muon-like'' subset are characterized by more uniform, 
track-like light deposition and are more easily reconstructed by existing reconstruction 
algorithms than ``cascade-like'' events. A reconstruction algorithm based on 
parameterization of time residuals from simulated muon bundles is used to reconstruct the 
zenith angle of the events (Fig. \ref{figzen}). Since most background muon bundles will 
come from a downgoing direction, while UHE neutrinos will come primarily from the 
horizontal direction \citep{kle99}, requiring that the zenith angle $>$ 85$^{\circ}$ 
(where a zenith angle of 90$^{\circ}$ is horizontal) reduces the background by a factor 
of 30.  The remaining background in the ``muon-like'' subset are misreconstructed events, 
since the actual flux close to the horizon is very small. A reconstruction based 
on the hit pattern of a Cherenkov cone for a minimally ionizing muon is applied 
to these events \citep{ahr04a}. Selecting only well-reconstructed events using the 
likelihood of this reconstruction (L$\mathrm{_{muon}}$) is sufficient to remove all 
background events in this subset. The value of this selection criteria was initially 
chosen to optimize the MRF for muon neutrinos with an E$^{-2}$ spectrum. However, by 
increasing the selection value slightly beyond the value which gave the minimum MRF, all 
background events were rejected with only a few percent drop in the sensitivity (Fig. 
\ref{figmrf}). Since the uncertainty in the cosmic ray spectrum is very large at these 
energies, the more stringent selection criterion was applied to correct for the fact that 
the MRF is optimized without uncertainties. 

The number of events at each selection level for experiment, background, and signal 
simulation for the ``muon-like'' subset are shown in Table \ref{tbl-passm}.


\section{Statistical and Systematic Uncertainties}\label{sec:unc}
Because there is no test beam which can be used to determine the absolute sensitivity of the 
AMANDA-II detector, calculations of sensitivity rely on simulation. The dominant sources 
of statistical and systematic uncertainty in this calculation are described below. The 
systematic uncertainties are assumed to have a flat distribution and are summed in 
quadrature separately for background and signal. The uncertainties have been included into 
the final limit using the method described in \citet{teg05}. 

\subsection{Uncertainties Due to Limited Simulation Statistics}
Due to computational requirements, background simulation statistics are somewhat limited. 
Ideally, one would scale the statistical uncertainty on zero events based on the 
simulation event weights in nearby non-zero bins.  However, the optimized background 
simulations used in this analysis have large variations in event weights approaching this 
region, making determination of this factor difficult.  Nevertheless, the statistical 
uncertainties near the edge of the distribution are on the order of the uncertainties for 
a simulation with a livetime equivalent to the data taking period, so no scaling factor 
is applied to the statistical uncertainty. A statistical uncertainty of 1.29, the 
1$\sigma$ Feldman-Cousins event upper limit on zero observed events \citep{fel98}, is 
assumed at the final selection level. Signal simulation has an average statistical 
uncertainty of 5\% for each neutrino flavor.

\subsection{Normalization of Cosmic Ray Flux}
The average energy of cosmic ray primaries at the penultimate selection level is 4.4 $\times$ 
10$^{7}$~GeV, which is considerably above the knee in the all-particle cosmic ray 
spectrum.
Numerous experiments have measured a large spread in the absolute normalization of the 
flux of cosmic rays at this energy (see \citet{kam07} for a recent review). Estimates of 
the uncertainty in the normalization of the cosmic ray flux range from 20\% \citep{hor03} 
to a factor of two \citep{pdg06}. This analysis uses the more conservative uncertainty of 
a factor of two.

\subsection{Cosmic Ray Composition}
There is considerable uncertainty in the cosmic ray composition above the knee 
\citep{pdg06}. We estimate the systematic uncertainty by considering two cases: 
proton-dominated composition and iron-dominated composition. The simulated background 
cosmic ray flux is approximated by separately treating proton and iron primaries 
combined in a total spectrum that becomes effectively iron-dominated above 10$^{7}$~GeV 
using the method described in \citet{gla99}. The iron-dominated spectrum yields a 30\% 
higher background event rate than the rate from a proton-dominated spectrum at the 
penultimate selection level. This value of 30\% is used as the uncertainty due to the 
cosmic ray composition.

\subsection{Detector Sensitivity}
The properties of the refrozen ice around each OM, the absolute sensitivity of 
individual OMs, and obscuration of OMs by nearby power cables can effect the detector 
sensitivity. This analysis uses the value obtained in \citet{ahr03a} where reasonable 
variations of these parameters in the simulation were found to cause a 15\% variation in 
the E$^{-2}$ signal and background passing rate.

\subsection{Implementation of Ice Properties}
As photons travel through the ice they are scattered and absorbed. The absorption and 
scattering lengths of the ice around the AMANDA-II detector have been measured very 
accurately using in situ light sources \citep{ack06}. Uncertainties are introduced due to 
the limited precision with which these parameters are included in the simulation. Varying 
the scattering and absorption lengths in the detector simulation by 10\% were found to 
cause a difference in number of expected signal events (for an E$^{-2}$ spectrum) of 34\% 
\citep{ack05}, which is used as a conservative estimate of the uncertainty due to 
implementation of ice properties. If too large of a deviation in background rate 
relative to the experimental rate was observed for a set of ice property parameters, the 
background rate was normalized to the experimental rate, and the signal rate was scaled 
accordingly. This was done to ensure that the variation in absorption and scattering 
lengths covered a reasonable range of ice properties.

\subsection{Neutrino Cross Section}
The uncertainty in the standard model neutrino cross section has been quantified recently 
\citep{anc06} taking into account the experimental uncertainties on the parton 
distribution functions measured at HERA \citep{che05}, as well as theoretical 
uncertainties in the effect of heavy quark masses on the parton distribution function 
evolution and on the calculation of the structure functions. The corresponding maximum 
variation in the number of expected signal events (for an E$^{-2}$ spectrum) is 10\%, in 
agreement with previous estimates \citep{ack05}.

\begin{figure}
\plotone{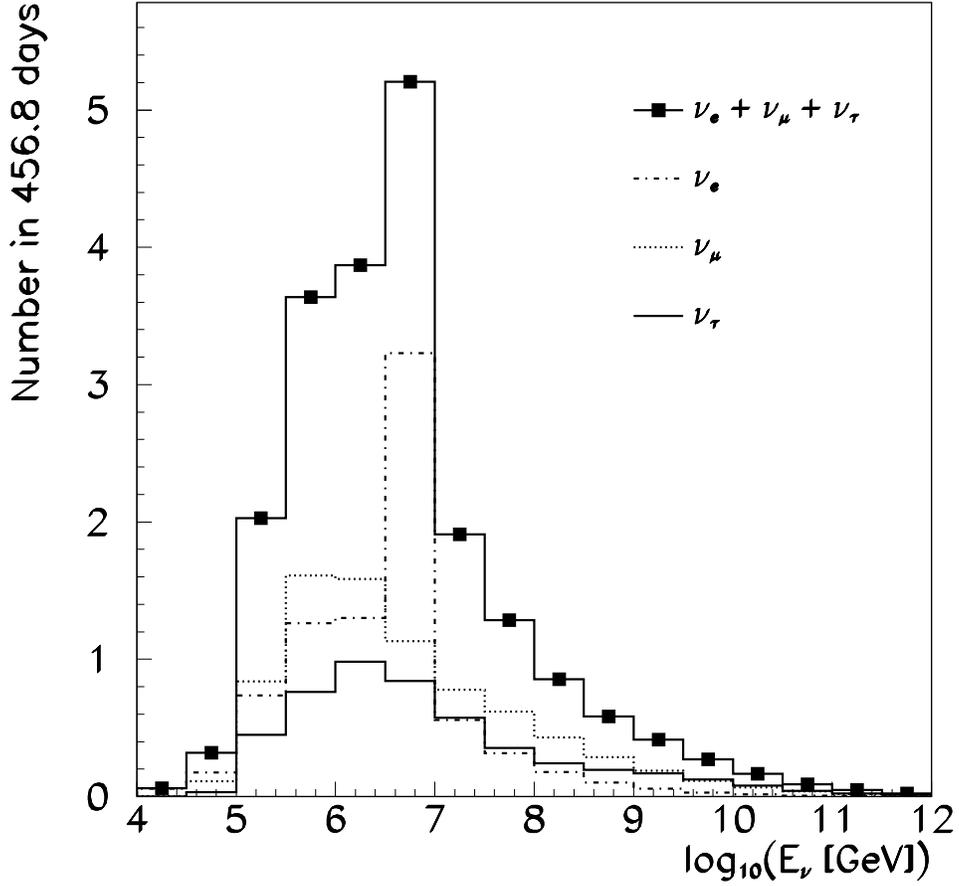}
\caption{Energy spectra of electron, muon, and tau neutrino signal events 
(d(N$_{\nu_e}$+N$_{\nu_\mu}$+N$_{\nu_\tau}$)/dE $=$ 10$^{-6}$~$\times$
E$^{-2}$~GeV $^{-1}$~cm$^{-2}$~s$^{-1}$~sr$^{-1}$) which pass all selection 
criteria. The peak in the electron neutrino spectrum just below 10$^{7}$~GeV is due 
to the Glashow resonance.\label{fig-spect}}
\end{figure}

Screening effects are expected to suppress the neutrino-nucleon cross section at energies 
in excess of 10$^{8}$~GeV (see e.g. \citet{kut03, ber07}). This has a negligible effect 
on the number of signal events expected for an E$^{-2}$ spectrum because the majority of 
signal is found below these energies (Fig. \ref{fig-spect}). Even if the suppression is as 
extreme as in the Colour Glass Condensate model \citep{hen05}, the event rate decreases 
by only 11\%.

\subsection{Differences in Simulated Distributions}
\begin{figure}
\plotone{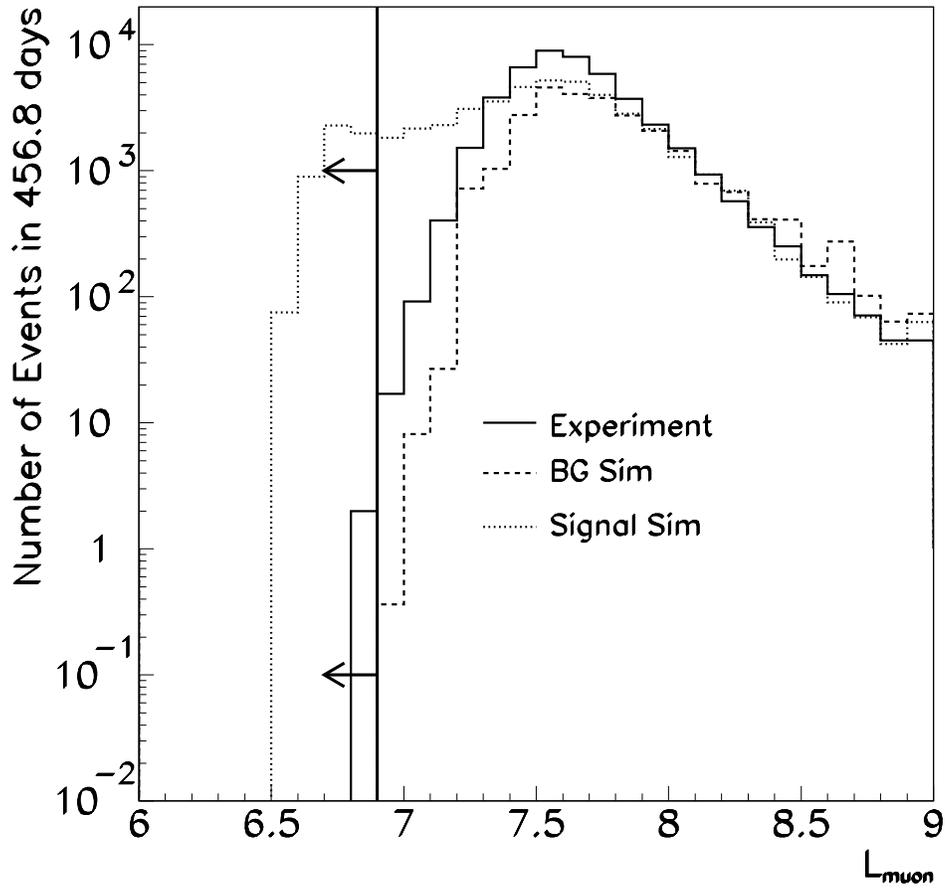}
\caption{L$_{\mathrm{muon}}$ distribution for the experiment, background, and 4.5 
$\times$ 10$^{-4}$ $\times$ E$^{-2}$ muon neutrino signal simulations (arbitrary 
normalization) in the ``muon-like'' subset after level three of this analysis. Two 
experimental events survive the final selection criteria of L$_{\mathrm{muon}}$ $<$ 6.9. 
\label{figfinal}}
\end{figure}

An examination of the L$_{\mathrm{muon}}$ distribution for the ``muon-like'' subset after 
level 3 of this analysis suggests the background simulation is shifted by one bin 
relative to the experiment (Fig. \ref{figfinal}). Shifting all simulation distributions to 
the left by one bin leads to better agreement between the background simulation and 
experimental distributions and an increase in 8\% in the number of expected signal events 
for an E$^{-2}$ spectrum.

\subsection{The Landau-Pomeranchuk-Migdal (LPM) Effect}
At ultra high-energies, the LPM effect suppresses the bremsstrahlung cross section for 
electrons and the pair-production cross section of photons created in a cascade by an 
electron neutrino \citep{lan53, mig57}. This lengthens the resultant shower produced by a 
factor that goes as $\sqrt{E}$. Above 10$^{8}$~GeV, the extended shower length becomes 
comparable to the spacing between OMs on a string \citep{kle04}. Additionally, as the LPM 
effect suppresses the bremsstrahlung and pair productions cross sections, photonuclear 
and electronuclear interactions begin to dominate which lead to the production of muons 
inside the electromagnetic cascade. Toy simulations were performed which superimposed a 
muon with an energy of 10$^{5}$~GeV onto a cascade with energy of 10$^{8}$~GeV. While the 
addition of the muon shifted the L$\mathrm{_{casc}}$ distribution 5\% towards higher 
(more ``muon-like'') values, the resulting events still passed all selection criteria 
indicating that the effects of muons created inside cascades are negligible.

The LPM effect is not included in the simulations of electron neutrinos, but it can be 
approximated by excluding all electron neutrinos with energies in excess of 10$^{8}$~GeV. 
This is an overestimation of the uncertainty introduced by the LPM effect, as extended 
showers may manifest as several separate showers which are likely to survive all selection 
criteria and the addition of low-energy muons is not expected to significantly alter the 
UHE cascade light deposition. Neglecting electron neutrinos with energies in excess of 
10$^{8}$~GeV reduces the number of expected signal events by 2\% for an E$^{-2}$ 
spectrum.

\subsection{Summary of Uncertainties}
The systematic errors are shown in Table \ref{tbl-unc}. Summing the systematic errors of 
the signal simulation in quadrature gives a systematic uncertainty of $\pm$39\%. 
Combining this with the statistical uncertainty of 5\% per neutrino flavor gives a total 
maximum uncertainty of 40\%. Following a similar method for the background simulation, the 
systematic uncertainty is +101\% / -60\%. Scaling the statistical uncertainty of the 
background simulation by the systematic uncertainty gives a maximum background expectation 
of fewer than 2.6 events for three years.

\section{Results}
After applying all selection 
criteria, no background events are expected for 456.8 days. Incorporating the statistical 
and systematic uncertainties, the background is expected to be found with a uniform prior 
probability between 0 and 2.6 events. A possible sensitivity calculation which 
incorporates these uncertainties can be generated by assuming a flat prior with a mean of 
1.3 events and a corresponding data expectation of 1 event. This gives a 90\% CL event 
upper limit of 3.5 \citep{teg05} and a sensitivity of 1.8 $\times$ 10$^{-7}$ GeV 
cm$^{-2}$ s$^{-1}$ sr$^{-1}$, with the central 90\% of the E$^{-2}$ signal found in the 
energy range 2 $\times$ 10$^{5}$~GeV to 10$^{9}$~GeV. Table \ref{tbl-num} shows the 
expected number of each flavor of UHE neutrino passing the final selection level for a 
10$^{-6}$ $\times$ E$^{-2}$ flux. The energy spectra of each flavor are shown in Fig. 
\ref{fig-spect}.

Two events are observed in the data sample at the final selection level (Fig. 
\ref{figfinal}), while fewer than 2.6 background events are expected which gives a  
90\% CL average event upper limit of 5.3. After applying all selection criteria, 20 events 
are expected for a 10$^{-6}$ $\times$ E$^{-2}$ all flavor flux (Table \ref{tbl-num}). The 
upper limit on the all-flavor neutrino flux (assuming a 1:1:1 
$\nu_e:\nu_{\mu}:\nu_{\tau}$ flavor ratio) is 
\begin{equation}
\mathrm{E^{2}\Phi_{90\% CL} \le 2.7 \times 10^{-7} GeV\ cm^{-2}\ s^{-1}\ sr^{-1},}
\end{equation}
\noindent
including systematic uncertainties, with the central 90\% of the E$^{-2}$ signal found 
between the energies of 2 $\times$ 10$^{5}$~GeV and 10$^{9}$~GeV. 

\begin{figure}
\plotone{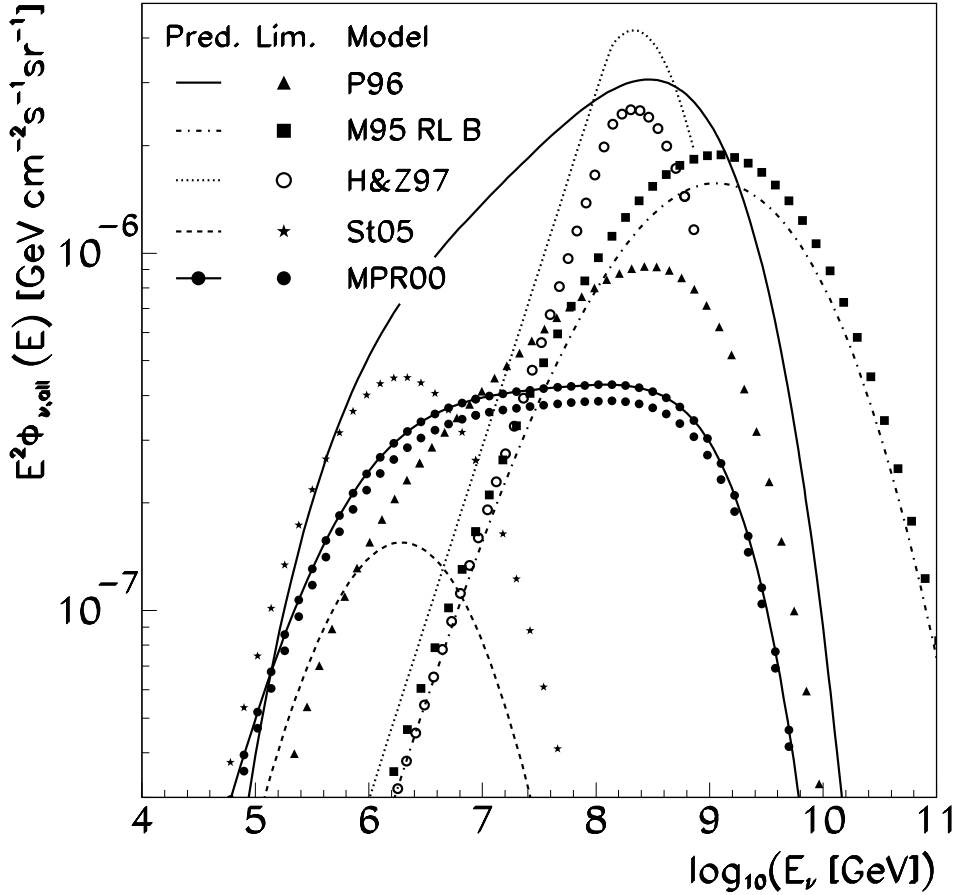}
\caption{AGN fluxes tested in this paper. Lines denote the model predictions and symbols 
denote the 90\% CL limits on the predictions derived by this analysis. The models rejected 
at the 90\% CL shown are: H\&Z97 from \citet{hal97}, P96 from \citet{pro96}, and MPR00 
from \citet{mpr00}. Also shown are models close to being rejected: M95 RL B from 
\citet{man95} and St05 from \citet{ste05}. See Table \ref{tbl-mrf} for exact 
numbers.\label{figagn}} 
\end{figure}

A number of theories which predict fluxes with non-E$^{-2}$ spectral shapes (Fig. 
\ref{figrep}) were also tested by reweighting the simulated signal. These include the 
hidden-core AGN model of \citet{ste92} which has been updated to reflect a better 
understanding of AGN emission \citep{ste05}, as well as AGN models in which neutrinos are 
accelerated in optically thin regions \citep{pro96,hal97,man95,mpr00}. Including 
uncertainties, this analysis restricts at a 90\% CL the AGN models from \citet{hal97} and 
\citet{mpr00}. Also the previously rejected \citep{ack05} models from \citet{pro96} and 
\citet{ste92} are rejected at the 90\% CL by this analysis (see Fig. \ref{figagn} and 
Table \ref{tbl-mrf}). The model by \citet{ste92} builds on a correlation between X-rays 
and neutrinos from AGNs. Other models using the same correlation give a similar 
normalization and violate current limits by an order of magnitude as well. As previously 
pointed out by \citet{bec07}, such a correlation can be excluded. 

While we do not directly exclude the flux from the \citet{ste05} hidden-core AGN model, 
it is possible to set limits on the parameters used in the model. In this model, the flux 
of neutrinos is normalized to the extragalactic MeV photon flux measured by COMPTEL with 
the assumption that the flux of photons from Seyfert galaxies is responsible for 10\% of 
this MeV background. If the neutrino flux scales linearly with the photon flux, then the 
maximum contribution of hidden-core AGNs, such as Seyfert galaxies, to the extragalactic 
MeV photon flux must be less than 29\%.

Fluxes of neutrinos from the decay of topological defects \citep{sig98} and the UHE 
fluxes required for the Z-bursts mechanism \citep{yos98,kal02a} peak at too high of an 
energy to be detected by this analysis. Neutrinos from the interaction of cosmic rays 
with cosmic microwave background photons are produced at too low of a flux for this 
analysis to detect (see Table \ref{tbl-mrf}).

\begin{figure}
\plotone{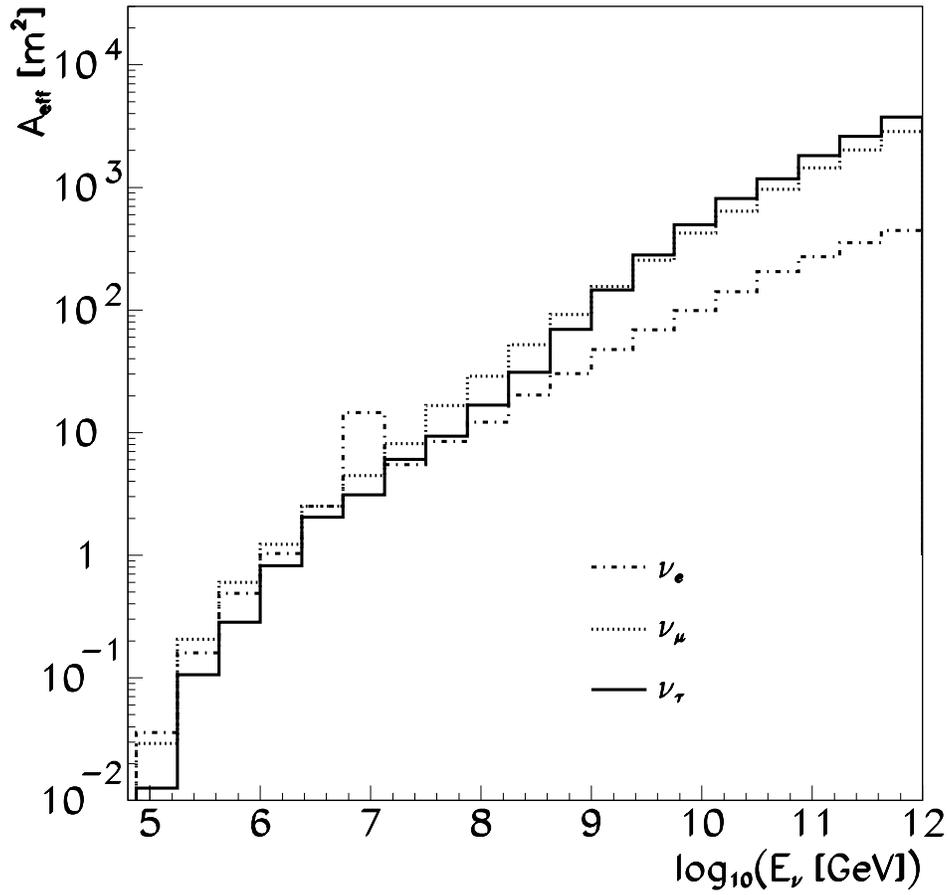}
\caption{Angle-averaged neutrino effective area at final selection level as a function of 
neutrino energy for electron, muon, and tau neutrinos. The peak in the electron neutrino 
effective area just below 10$^{7}$~GeV is due to the Glashow resonance.\label{figaeff}} 
\end{figure}

The number of expected events of a given flavor ($\nu$ and $\overline{\nu}$) for 
spectra not tested in this paper can be calculated using the formula
\begin{equation}
N_{signal}=T\int{dE_{\nu}d\Omega{\Phi_{\nu}(E_{\nu})A_{\it eff}^{\nu}(E_{\nu}})},
\end{equation}
\noindent
where T is the total livetime (456.8 days), A$_{\it eff}^{\nu}$ is the angle averaged 
neutrino effective area (Fig. \ref{figaeff}), and $\Phi_{\nu}$ is the flux at the Earth's 
surface.

\section{Conclusion}
The diffuse neutrino flux limit for a 1:1:1 $\nu_e:\nu_{\mu}:\nu_{\tau}$ flavor ratio set 
by this analysis of 
\begin{equation}
\mathrm{E^{2}\Phi_{90\% CL} \le 2.7 \times 10^{-7} GeV\ cm^{-2}\ s^{-1}\ sr^{-1},}
\end{equation}
is the most stringent to date above 10$^{5}$~GeV. A number of models for neutrino 
production have been rejected (see Table \ref{tbl-mrf} for a full list). AMANDA-II 
hardware upgrades which were completed in 2003 should lead to an improvement of the 
sensitivity at ultra high-energies \citep{sil05}. Additionally, AMANDA-II is now 
surrounded by the next-generation IceCube detector which is currently under construction. 
The sensitivity to UHE muon neutrinos for 1 year is expected to increase by roughly an 
order of magnitude as the IceCube detector approaches its final size of 1~km$^{3}$ 
\citep{ahr04b}.

\acknowledgments
We acknowledge the support from the following agencies: National Science 
Foundation-Office of Polar Program; National Science Foundation-Physics Division; 
University of Wisconsin Alumni Research Foundation; Department of Energy, and National 
Energy Research Scientific Computing Center (supported by the Office of Energy Research 
of the Department of Energy); the NSF-supported TeraGrid system at the San Diego 
Supercomputer Center (SDSC); the National Center for Supercomputing Applications 
(NCSA); Swedish Research Council; Swedish Polar Research Secretariat; Knut and Alice 
Wallenberg Foundation, Sweden; German Ministry for Education and Research; Deutsche 
Forschungsgemeinschaft (DFG), Germany; Fund for Scientific Research (FNRS-FWO); Flanders 
Institute to encourage scientific and technological research in industry (IWT); Belgian 
Federal Office for Scientific, Technical and Cultural affairs (OSTC); the Netherlands 
Organisation for Scientific Research (NWO); M. Ribordy acknowledges the support of the 
SNF (Switzerland); A. Kappes and J. D. Zornoza acknowledges the Marie Curie OIF Program; 
L. Gerhardt acknowledges the support of the University of California, Irvine MPC 
Computational Cluster and Achievement Rewards for College Scientists (ARCS).

\clearpage

\begin{table}
\begin{center}
\caption{Selection criteria.\label{tbl-cuts}}
\begin{tabular}{cll}
\tableline\tableline
Level & \multicolumn{2}{c}{Selection Criteria}\\
\tableline
0 &\multicolumn{2}{c}{Hit Cleaning and Retriggering}\\
1 &\multicolumn{2}{c}{F1H $<$ 0.72}\\
 &\multicolumn{2}{c}{NHITS $>$ 140}\\
2 &\multicolumn{2}{c}{F1H $<$ 0.53}\\
\tableline
 &``Cascade-like'' &``Muon-like''\\
\tableline
3 &L$_\mathrm{{casc}}$ $<$ 7& L$_\mathrm{{casc}}$ $\ge$ 7\\
 &Neural Net $>$ 0.93 &Zenith Angle $>$ 85\\
4 &F4H $<$ 0.1 &L$_{\mathrm{muon}}$ $<$ 6.9\\
5 &F1H$_{\mathrm{ELEC}}$ $<$ 0.56 & -\\
6 &L$_{60}$ $<$ 6.6 & -\\
\tableline
\end{tabular}
\end{center}
\end{table}

\clearpage
\begin{table}
\begin{center}
\caption{Number of experimental, simulated background, and simulated signal events in the 
``cascade-like'' subset at each selection level for 456.8 days.\label{tbl-passc}} 
\begin{tabular}{ccccccc}
\tableline\tableline
Level & Data & BG Simulation & Signal Simulation\\
\tableline
0 &2.7 $\times 10^{9}$ & 1.8$^{+1.8}_{-1.1}$ $\times 10^{9}$ & 621.7\\
1 &3.9 $\times 10^{7}$ & 3.1$^{+3.1}_{-1.8}$ $\times 10^{7}$ & 270.8\\
2 &1.7 $\times 10^{4}$ &1.4$^{+1.4}_{-0.8}$ $\times 10^{4}$ &89.2\\
3 &155 & 62$^{+63}_{-37}$ &32.0\\
4 &151 & 61$^{+62}_{-37}$ &31.0\\
5 &46 & 32$^{+32}_{-19}$ &27.1\\
6 &0 &0$^{+2.6}$ &16.0\\
\tableline
\end{tabular}
\tablecomments{{Levels 0 and 1 show combined numbers for both ``muon-like'' and 
``cascade-like'' subsets. Signal is shown with a low energy threshold of 10$^{4}$~GeV for 
a neutrino spectrum of dN/dE $=$ 10$^{-6}$~$\times$ 
E$^{-2}$~GeV$^{-1}$~cm$^{-2}$~s$^{-1}$~sr$^{-1}$, with an assumed 1:1:1 
$\nu_e:\nu_{\mu}:\nu_{\tau}$ flavor ratio. Values at selection level 0 and 1 for data and 
background simulation are extrapolated from the 2000 datasets. The background simulation 
is shown with systematic and statistical uncertainties described in Section 
\ref{sec:unc}. The number of ``muon-like'' events are shown in Table \ref{tbl-passm}.}}
\end{center}
\end{table}

\clearpage
\begin{table}
\begin{center}
\caption{Number of experimental, simulated background, and simulated signal events in the 
``muon-like'' subset at each selection level for 456.8 days.\label{tbl-passm}} 
\begin{tabular}{ccccccc}
\tableline\tableline
Level & Data & BG Simulation & Signal Simulation\\
\tableline
0 &2.7 $\times 10^{9}$ & 1.8$^{+1.8}_{-1.1}$ $\times 10^{9}$ & 621.7\\
1 &3.9 $\times 10^{7}$ & 3.1$^{+3.1}_{-1.8}$ $\times 10^{7}$ & 270.8\\
2 &1.4 $\times 10^{6}$ &9.0$^{+9.1}_{-5.4}$ $\times 10^{5}$ &85.2\\
3 &4.6 $\times 10^{4}$ &2.7$^{+2.7}_{-1.6}$ $\times 10^{4}$ &57.9\\
4 &2 &0$^{+2.6}$ &4.0\\
\tableline
\end{tabular}
\tablecomments{{Levels 0 and 1 show combined numbers for both ``muon-like'' and 
``cascade-like'' subsets. Signal is shown with a low energy threshold of 10$^{4}$~GeV for 
a neutrino spectrum of dN/dE $=$ 10$^{-6}$~$\times$ 
E$^{-2}$~GeV$^{-1}$~cm$^{-2}$~s$^{-1}$~sr$^{-1}$, with an assumed 1:1:1 
$\nu_e:\nu_{\mu}:\nu_{\tau}$ flavor ratio. Values at selection level 0 and 1 for data and 
background simulation are extrapolated from the 2000 datasets. The background simulation 
is shown with systematic and statistical uncertainties described in Section 
\ref{sec:unc}. The number of ``cascade-like'' events are shown in Table \ref{tbl-passc}.}}
\end{center}
\end{table}

\clearpage    
\begin{table} 
\begin{center}
\caption{Simulation Uncertainties\label{tbl-unc}}
\begin{tabular}{lrr}
\tableline\tableline
Source & BG Sim & Sig Sim \\
\tableline
Cosmic Ray Normalization & +100\% / -50\%&-\\
Cosmic Ray Composition & -30\%&-\\
Detector Sensitivity & $\pm$15\%& $\pm$15\%\\
Ice Properties &-& $\pm$34\%\\
Neutrino Cross Section &- &$\pm$10\%\\
Simulation Distribution &-&+8\%\\
LPM Effect &-&-2\%\\
\tableline
Total & +101\% / -60\%& +39\% / -39\%\\
\tableline
\end{tabular}
\end{center}
\end{table}

\clearpage
\begin{table}
\begin{center}
\caption{Number of simulated neutrino events in the ``cascade-like'' and ``muon-like'' 
subsets passing all selection criteria for three years for a neutrino spectrum of 
d(N$_{\nu_e}$+N$_{\nu_\mu}$+N$_{\nu_\tau}$)/dE $=$ 10$^{-6}$~$\times$ 
E$^{-2}$~GeV $^{-1}$~cm$^{-2}$~s$^{-1}$~sr$^{-1}$.\label{tbl-num}}
\begin{tabular}{lccc}
\tableline\tableline
Neutrino Flavor & ``Cascade-like''& ``Muon-like''& Total\\
\tableline
Electron &7.7 &0.1 &7.8\\
Muon &3.9 &3.6 &7.5\\
Tau &4.4 &0.3 &4.7\\
All Flavors &16.0 &4.0 &20.0\\
\tableline
\end{tabular}
\end{center}
\end{table}

\clearpage
\begin{table}
\begin{center}
\caption{Flux models, the number of neutrinos of all flavors expected at the Earth at the 
final selection level, and the MRFs for 456.8 days of livetime.\label{tbl-mrf}}
\begin{tabular}{lccc}
\tableline\tableline

Model & $\nu_{all}$ & MRF & Reference\\
\tableline
AGN\tablenotemark{a} &20.6 &0.3 &\citep{pro96}\\
AGN\tablenotemark{a} &17.4 &0.3 &\citep{ste92}\\
AGN\tablenotemark{a} &8.8 &0.6 &\citep{hal97}\\
AGN\tablenotemark{a} &5.9 &0.9 &\citep{mpr00}\\
AGN RL B\tablenotemark{a} &4.5 &1.2 &\citep{man95}\\
Z-burst &2.0 &2.7 &\citep{kal02a}\\
AGN &1.8 &2.9 &\citep{ste05}\\
GZK $\nu$ norm AGASA\tablenotemark{b} &1.8 &2.9 &\citep{ahl05}\\
GZK $\nu$ mono-energetic &1.2 &4.4 &\citep{kal02b}\\
GZK $\nu$ a$=$2 &1.1 &4.8 &\citep{kal02b}\\
GZK $\nu$ norm HiRes\tablenotemark{b} &1.0 &5.3 &\citep{ahl05}\\
TD &0.9 &5.9 &\citep{sig98}\\
AGN RL A\tablenotemark{a} &0.3 &18.0 &\citep{man95}\\
Z-burst &0.1 &53.0 &\citep{yos98}\\
GZK $\nu$ &0.06 &88.0 &\citep{eng01}\\
\tableline
\end{tabular}
\tablenotetext{a}{{These values have been divided by two to account for neutrino oscillation 
from a source with an initial 1:2:0 $\nu_{e}$:$\nu_{\mu}$:$\nu_{\tau}$ flux.}}
\tablenotetext{b}{{Lower energy threshold of 10$^{7}$~GeV applied.}}
\tablecomments{{A MRF of less than one indicates that the model is excluded with 90\% 
confidence.}}
\end{center}
\end{table}

\clearpage
\thispagestyle{empty}

\clearpage


\begin{thebibliography}{}
\bibitem[Achterberg et al.(2007)]{ach07} Achterberg, A. et al., 2007 Phys. Rev. D, in press (ArXiv:0705.1315).
\bibitem[Ackermann et al.(2004)]{ack04} Ackermann, M. et al., 2004, Astroparticle Physics, 22, 127.
\bibitem[Ackermann et al.(2005)]{ack05} Ackermann, M. et al., 2005, Astroparticle Physics, 22, 339.
\bibitem[Ackermann et al.(2006)]{ack06} Ackermann, M. et al., 2006, Journal of Geophysical Research, 3, D13203.
\bibitem[Ahlers et al.(2005)]{ahl05} Ahlers, M. et al., 2005, Phys. Rev. D, 72, 023001.
\bibitem[Ahrens et al.(2003a)]{ahr03a} Ahrens, J. et al., 2003, Astrophys. Journal, 583, 1040.
\bibitem[Ahrens et al.(2003b)]{ahr03b} Ahrens, J. et al., 2003, Phys. Rev. Lett., 90, 251101.
\bibitem[Ahrens et al.(2003c)]{ahr03c} Ahrens, J. et al., 2003, Phys. Rev. D, 67, 012003.
\bibitem[Ahrens et al.(2004a)]{ahr04a} Ahrens, J. et al., 2004, Nuclear Instruments and Methods A, 524, 169.
\bibitem[Ahrens et al.(2004b)]{ahr04b} Ahrens, J. et al., 2004, Astroparticle Physics, 20, 507.
\bibitem[Anchordoqui et al.(2006)]{anc06} Anchordoqui, L., Cooper-Sakar, A., Hooper D., and, Sakar, S., 2006, Phys, Rev. D, 74, 043008.
\bibitem[Aynutdinov et al.(2006)]{ayn06} Aynutdinov, V. et al., 2006, Astroparticle Physics, 140.
\bibitem[Bahcall \& Waxman(1998)]{bah98} Bahcall, J. and Waxman, E., 1998, Phys. Rev. D, 59, 023002.
\bibitem[Barwick et al.(2006)]{bar06} Barwick, S. et al., 2006, Phys. Rev. Lett., 96, 171101.
\bibitem[Becker et al.(2007)]{bec07} Becker, J.~K., {Gro{\ss}}, A., {M\"unich}, K., {Dreyer}, J., {Rhode}, W. and {Biermann}, P.~L., 2007, Astroparticle Physics, in press (astro-ph/0607427).
\bibitem[Berger et al.(2007)]{ber07} Berger, E., Block, M., McKay, D., and Tan, C., 2007, preprint (hep-ph/0708.1960v1).
\bibitem[Chekanov et al.(2005)]{che05} Chekanov, S. et al., 2005, European Physical Journal C, 42, 1.
\bibitem[Chirkin \& Rhode(2004)]{chi04} Chirkin, D. and Rhode, W., 2004, preprint (hep-ph/0407075).
\bibitem[Engel et al.(2001)]{eng01} Engel, R., Seckel, D., and Stanev T., 2001, Phys. Rev. D, 64, 093010. Curve shown for $\Lambda$$=$0.7, taken from ftp://ftp.bartol.udel.edu/seckel/ess-gzk/flux\_n3\_8\_flat\_om0p3.txt.
\bibitem[Feldman \& Cousins(1998)]{fel98} Feldman, G. and Cousins, F., 1998, Phys. Rev. D, 57, 3873.
\bibitem[Gandhi et al.(1998)]{gan98} Gandhi, R., Quigg, C., Reno, M., and Sarcevic, I., 1998, Phys. Rev. D, 58, 093009.
\bibitem[Gerhardt(2005)]{ger05} Gerhardt, L., 2005, Proc. 29th Int. Cosmic Ray Conf., Pune, India, 111 (astro-ph/0509330).
\bibitem[Gerhardt(2007)]{ger06} Gerhardt, L., 2007, Proc. 14th Int. Conf. on Supersymmetry and the Unification of Fundamental Interactions, AIP Conf. Proc. 903, 622.
\bibitem[Glasstetter et al.(1999)]{gla99} Glasstetter, R. et al., 1999, Proc. 26th Int. Cosmic Ray Conf., Utah, USA, 1, 222.
\bibitem[Halzen \& Zas(1997)]{hal97} Halzen, F. and Zas, E., 1997, Astrophys. Journal, 488, 669.
\bibitem[Hamamatsu(1999)]{ham99} Hamamatsu, 1999, Photomultiplier Tubes, Basics and Applications, second ed.
\bibitem[Heck(1999)]{hec99} Heck, D., 1999, DESY-PROC-1999-01, 227.
\bibitem[Henley \& Jalilian-Marian(2005)]{hen05} Henley, E. and Jalilian-Marian, J., 2005, preprint (hep-ph/0512220v1).
\bibitem[Hill \& Rawlins(2003)]{hil03} Hill, G. and Rawlins, K., 2003, Astroparticle Physics, 19, 393.
\bibitem[H\"orandel(2003)]{hor03} H\"orandel, J., 2003, Astroparticle Physics, 19, 193.
\bibitem[Hundertmark(1998)]{hun98} Hundertmark, S., 1998, Proc. 1$^{st}$ Workshop Methodical Aspects of Underwater/Ice Neutrino Telescopes, Zeuthen, Germany.
\bibitem[Kalashev et al.(2002a)]{kal02a} Kalashev, O. et al., 2002, Phys. Rev. D 65, 103003.
\bibitem[Kalashev et al.(2002b)]{kal02b} Kalashev, O. et al., 2002, Phys. Rev. D 66, 063004.
\bibitem[Kampert(2007)]{kam07} Kampert, K.-H., 2007, Nuclear Physics B (Proc. Suppl.), 165, 294.
\bibitem[Kashti \& Waxman(2005)]{kas05} Kashti, T. and Waxman, E. Phys. Rev. Lett. 95, 181101.
\bibitem[Klein \& Mann(1999)]{kle99}  Klein, J. and Mann, A., 1999, Astroparticle Physics 10, 321.
\bibitem[Klein(2004)]{kle04} Klein, S., 2004, preprint (astro-ph/0412546v1).
\bibitem[Kowalski \& Gazizov(2005)]{kow05} Kowalski, M. and Gazizov, A., 2005, Computer Physics Communications, 172, 203. 
\bibitem[Kravchenko et al.(2006)]{kra06} Kravchenko, I. et al., 2006, Phys. Rev. D, 73, 082002.
\bibitem[Kutak \& Kwieci\`nski(2003)]{kut03} Kutak, K. and Kwieci\`nski, J., 2003, preprint (hep-ph/0303209v4).
\bibitem[Landau \& Pomeranchuk(1953)]{lan53} Landau, L. and. Pomeranchuk, I., 1953, Dok. Akad. Nauk SSSR, 92, 535.
\bibitem[Lipari(1993)]{lip93} Lipari, P., 1993, Astroparticle Physics, 1, 195. 
\bibitem[Mannheim(1995)]{man95} Mannheim, K., 1995, Astroparticle Physics, 3, 295.
\bibitem[Mannheim et al.(2000)]{mpr00} Mannheim, K., Protheroe, R. J., and Rachen, J., 2000, Phys. Rev. D, 63, 023003.
\bibitem[Migdal(1957)]{mig57} Migdal, A., 1957, JETP, 5, 527.
\bibitem[Particle Data Group(2006)]{pdg06} Particle Data Group, 2006, Journal of Phys. G. 33, 1.
\bibitem[Pohl(2004)]{poh04} Pohl A.C., 2004, Licenciate thesis, Kalmar University.
\bibitem[Protheroe(1996)]{pro96} Protheroe, R., 1996, preprint (astro-ph/9607165).
\bibitem[Silvestri(2005)]{sil05} Silvestri, A., 2005, Proc. 29th Int. Cosmic Ray Conf., Pune, India, 431.
\bibitem[Stecker(2005)]{ste05} Stecker, F.,2005, Phys. Rev. D, 72, 107301.
\bibitem[Stecker et al.(1992)]{ste92} Stecker, F., Done, C., Salamon, M., and Sommers, P., 1992, Phys. Rev. Lett., 69, 2738.
\bibitem[Sigl et al.(1998)]{sig98} Sigl, G., Lee, S., Bhattacharjee, P., and Yoshida, S., 1998, Phys. Rev. D, 59, 043504.
\bibitem[Tegenfeldt \& Conrad(2005)]{teg05} Tegenfeldt, F. and Conrad, J., 2005, Nuclear Instruments and Methods in Physics Research A, 539, 407.
\bibitem[Wiebel-Sooth et al.(1999)]{wie99} Wiebel-Sooth, B., Biermann, P., and Meyer, H., 1999, col. VI/3c, Springer Verlag, 37.
\bibitem[Yoshida et al.(1998)]{yos98} Yoshida, S., Sigl, G., and Lee S., 1998, Phys. Rev. Lett., 81, 5505.
\bibitem[Zas et al.(1993)]{zas93} Zas, E., Halzen, F., and V\'{a}zquez, R., 1993, Astroparticle Physics, 1, 297.
\end{thebibliography}
\end{document}